\documentclass[journal=jpclcd,manuscript=article]{achemso}
\usepackage[version=3]{mhchem} 
\usepackage{braket}
\usepackage{subcaption}
\usepackage{xcolor}
\usepackage{amsmath}
\usepackage{siunitx}
\usepackage{soul}

\definecolor{Burgundy}{RGB}{144,0,32}

\author{Brieuc Le D\'e}
\author{Simon Huppert}
\affiliation[INSP]
{Sorbonne Universit\'e, CNRS, Institut des NanoSciences de Paris, 4 place Jussieu, 75005 Paris, France}%
\author{Riccardo Spezia}
\affiliation[LCT]
{Sorbonne Universit\'e, CNRS, Laboratoire de Chimie Th\'eorique, 4 place Jussieu, 75005 Paris, France}%
\author{Alex W. Chin}
\email{alex.chin@insp.jussieu.fr}
\affiliation[INSP]
{Sorbonne Universit\'e, CNRS, Institut des NanoSciences de Paris, 4 place Jussieu, 75005 Paris, France}%

\title[HBQ/HBT wave packet study]
{ Impact and Interplay of Quantum Coherence and Dissipative Dynamics for Isotope Effects in Excited-State Intramolecular Proton Transfer }

\abbreviations{Isotope effect ESIPT}
\begin{document}

\begin{tocentry}
      \includegraphics[width=1.0\linewidth]{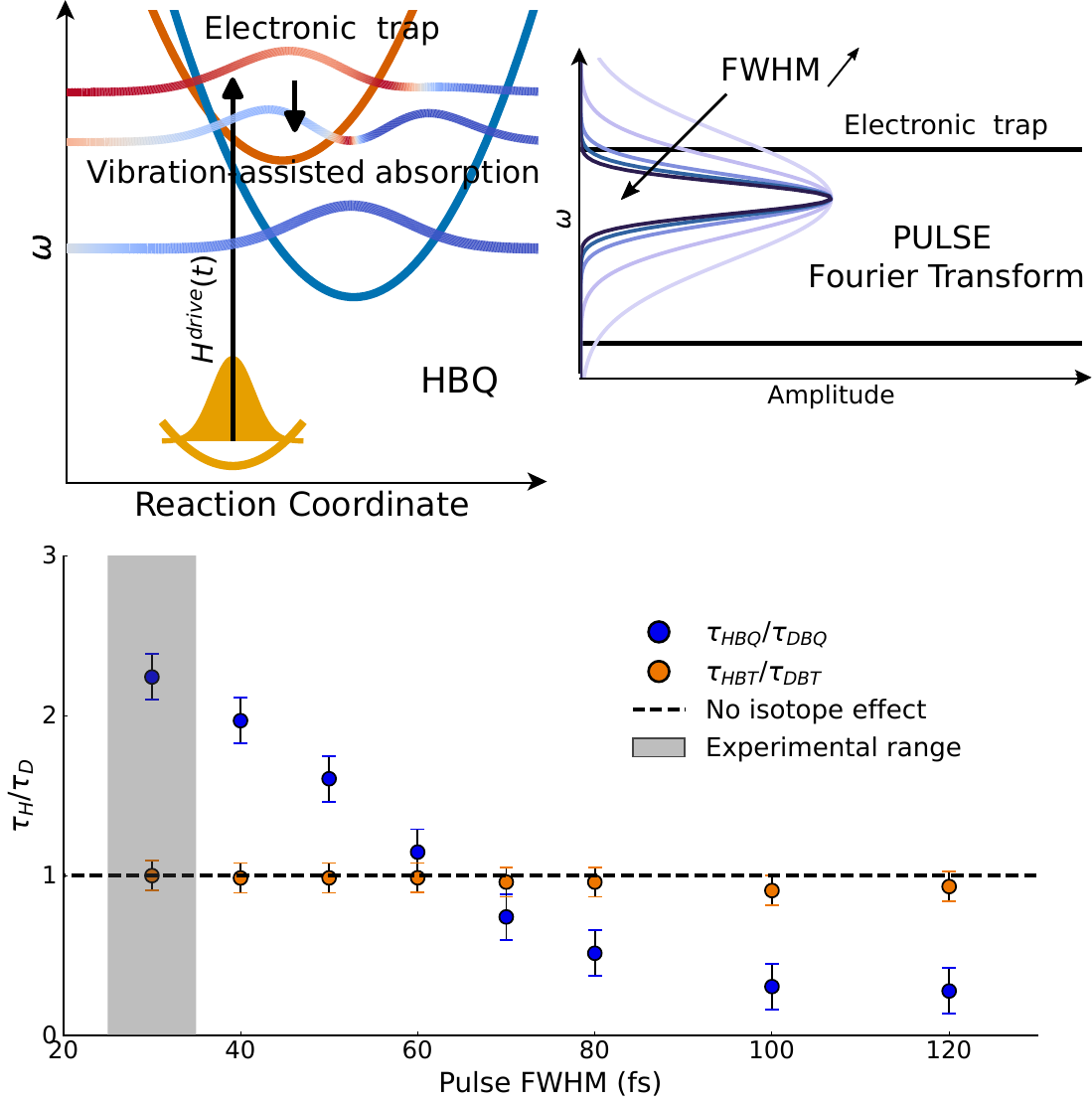}
\end{tocentry}

\begin{abstract}
\noindent 
The quantum dynamics of Excited State Intramolecular Proton Transfer (ESIPT) is studied using a non-Markovian open quantum system perspective. Models of 2-(2'-hydroxyphenyl) benzothiazole (HBT) and 10-hydroxybenzo[h]quinoline (HBQ) are adapted from Zhang \textit{et al.} (ACS Phys. Chem. Au, 3, 107–118 (2023)) and simulated via the numerically exact TEDOPA Matrix Product State formalism, using a newly developed framework for continuous degrees of freedom subject to dissipation. The quantum treatment of the proton wave packet shows a counter-intuitive kinetic isotope effect, with strong isotope dependence for the barrierless potential surface of HBQ and no isotope effect in the double-well energy landscape of the HBT, in accordance with experimental results. Strikingly, for HBQ we find that changing laser pulse durations can even reverse the isotope effect on the proton transfer rate, revealing the role of vibration-assisted absorption in ESIPT. This study highlights the often neglected effect of excitation conditions on ESIPT, as well as the role of entangled, vibrationally assisted absorption processes that can be directly visualised in our multidimensional treatment of the full electro-vibronic-environment wave function.
\end{abstract}

Excited-State Intramolecular Proton Transfer (ESIPT) is an ubiquitous tautomeric process with timescales ranging from few tens of femtoseconds \cite{AmeerBeg2001} up to picoseconds \cite{Alarcos2015}, and is accompanied by a modification of molecular electronic structures that can trigger a wide range of light-driven catalytic processes, including critical biosynthetic processes such as chlorophyll production \cite{heyes2017excited}.  ESIPT is often characterised by properties such as strikingly large Stokes shifts ($100$ - $250~$nm) \cite{Zhang2014,Shen2023} between the differently protonated excited-states, and has been proposed for applications including dual-fluorescence biological markers \cite{Gu2023,lama_unraveling_2022,Li2021,Sedgwick2018}, pH-dependent optical sensors \cite{Kwon2011,chen_excitedstate_2021,Sedgwick2018,Demchenko2023}, intrinsic white-light OLEDS \cite{joshi_excited-state_2021} and organic lasing \cite{yan2020organic}.

Understanding of the hydrogen/deuterium Kinetic Isotope Effect (KIE) provides deep insight into the mechanisms of ESIPT, and in particular the importance of the proton's intrinsic quantum dynamics during the photoreaction. Criticality, a number of possible scenarios have been proposed for ultrafast ESIPT in which the role of the proton's quantum nature is more or less `active', and, in principle, these can be distinguished by measurements of KIE \cite{lee_active_2013}. For example, proton tunnelling between vibronic states of electronically diabatic potentials would be expected to give a large KIE \cite{layfield_hydrogen_2014}, whereas skeletal (vibrational) deformations induced by quasi-impulsive Franck-Condon excitation could carry a `passive'  (classical) proton to the ESIPT transition state without any noticeable KIE \cite{Schriever2011}. For many systems, the mechanism seems to lie somewhere in between, and the exact proton transfer mechanism(s) are still a matter of active debate \cite{barbatti_ultrafast_2009,lee_active_2013,Kim2020,picconi_nonadiabatic_2021,Luber2013}. Recently, a range of advanced numerical descriptions of quantum and semi-classical ESIPT proton dynamics have revealed the importance of many different ultrafast processes in this complex phenomenon, such as non-adiabatic coupling, solvent interactions, vibrational energy redistribution (relaxation) and nonlinear coupling between vibration normal modes and/or the proton coordinate \cite{Hazra2010,HammesSchiffer2012,HammesSchiffer2008_KIEtheory,goyal_tuning_2017}. 

In this letter, we study the hydrogen/deuterium KIE using two models: the first one represents 2-(2'-hydroxyphenyl) benzothiazole (HBT) and its isotope (DBT), which are described by a barriered double-well energy landscape PES, while the second describes 10-hydroxybenzo[h]quinoline (HBQ) and its isotope (DBQ), which perform ESIPT on a barrierless PES. Our description is based on the phenomenological models recently developed by Zhang \textit{et al.}\cite{scholes_ref_2023} that reproduce the surprising experimental results that no KIE is observed for the slower ESIPT in the (double well) HBT (DBT) system, whereas a clear KIE appears in the barrierless case of HBQ (DBQ).  The study of Zhang \textit{et al.} points to an `active' role of the proton d.o.f. in both cases, with a key role played by non-adiabatic effects in their treatment of the excited state dynamics.  Importantly for this work, Zhang \textit{et al}. also emphasised the synergistic role of the dissipative environment in the anomalous KIEs, motivating the present work which uses the Time-Evolving Density with Orthogonal
Polynomials (TEDOPA) algorithm to explore ESIPT as an open quantum system problem \cite{mpsdynamicsjl_2024,dunnett2021efficient}. TEDOPA has been widely applied in diverse systems across quantum physics \cite{prior2013quantum} and -- more recently -- ultrafast photodynamics \cite{del2018tensor,dunnett2021influence,hunter2024environmentally}, and exploits the computational efficiency of Tensor Networks for describing the many body dynamics of large entangled wave functions \cite{garcia-ripoll_time_2006,chin_exact_2010,orus_practical_2014,paeckel_time-evolution_2019,cirac_matrix_2021,banuls_tensor_2023}. 

With access to the complete many body wave function of the electronic-proton-environment system, we show that unusual KIE effects can arise due to the interplay of dissipation and coherent (non-adiabatic) vibronic dynamics, highlighting the role of isotopic substitutions in controlling the eigenstate structure that can promote or suppress rapid ESIPT via synergistic dissipative effects. In particular, we show that even 
when the first adiabatic singlet excited state (S$_1$) presents a fast, barrierless route-to-product, non-adiabatic effects can actually impede ESIPT if the diabatic reactant/product states are energetically misaligned. A consequence is the very strong (non-monotonic) dependence of ESIPT on excitation conditions, which we observe and rationalise through interrogation of the complete TEDOPA  wave functions. 

\begin{figure}[ht!]
    \centering
    \includegraphics[scale=0.55]{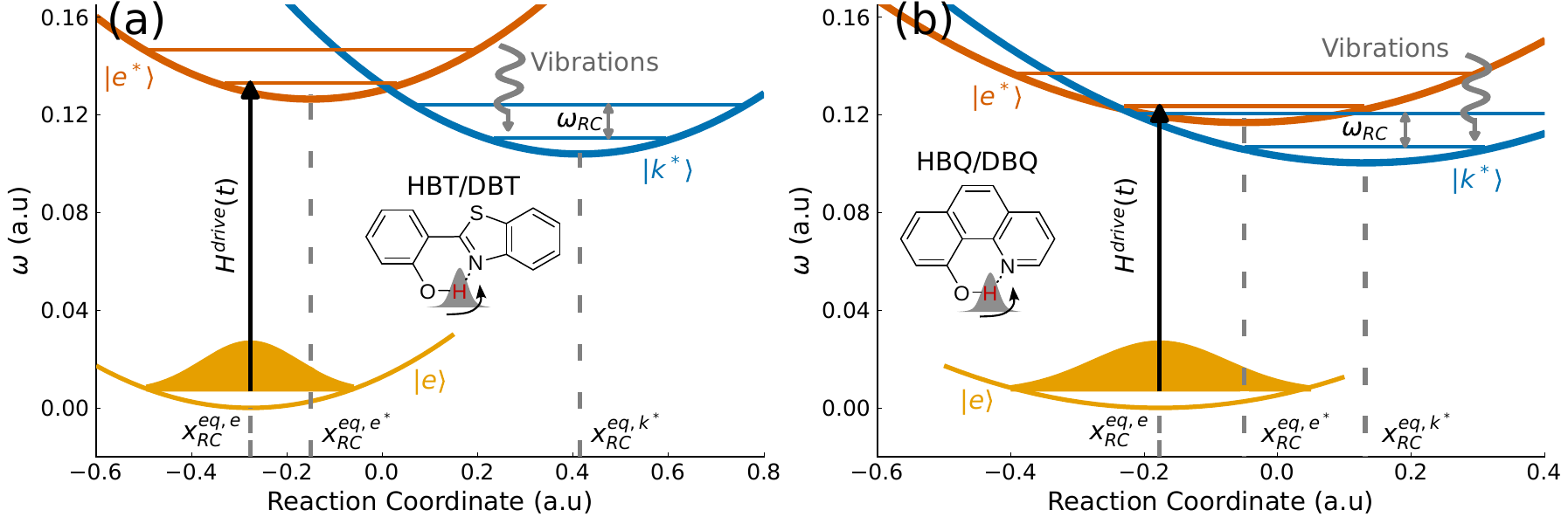}
    \caption{Model and dynamics parameters of (a) HBT and DBT (b) HBQ and DBQ.  The proton transfer in HBT and HBQ structures is also illustrated. The mass and frequency of the reaction coordinate oscillator are changed according to the studied isotope.}
    \label{fig:figPES}
\end{figure}

Figure~\ref{fig:figPES} illustrates the two models where the 1D intramolecular reaction coordinate (RC) is defined as the reaction path for the proton to move from the oxygen to the nitrogen. Following Zhang \textit{et al.},  our model includes three relevant electronic states: the ground state (enol) $\ket{e}$, the excited enol structure $\ket{e^*}$ and the excited keto $\ket{k^*}$ that results from the change of electronic bonding upon proton transfer to the nitrogen. 

The proton transfer reaction coordinate is described by a quantum oscillator with a characteristic frequency $ \omega_\text{RC}$ and mass $m_{\text{RC}}$, with standard (bosonic) annihilation (creation) operators $a^{\left( \dagger \right)}$. Employing atomic units, the main system can be described by the Hamiltonian $H_\text{S}$:
\begin{subequations}
\renewcommand{\theequation}{\theparentequation.\arabic{equation}}
   \begin{align}
          &H_\text{S} = H^\text{elec} + H^\text{RC} + H_\text{int}^\text{elec-RC} \\
             &H^\text{elec} = \omega_{e}  \ket{e}\bra{e} + \omega_{e^*}  \ket{e^*}\bra{e^*}  + \omega_{k^*} \ket{k^*}\bra{k^*}     + \Delta \Big( \ket{e^*}\bra{k^*} + \ket{k^*}\bra{e^*}\Big)\\
             &H^\text{RC} = \omega_\text{RC} \left( a^{\dagger}a + \frac{1}{2} \right)  \\
            &H_\text{int}^\text{elec-RC} = \Bigl[ \ket{e}\bra{e} y_{e} + \ket{e^*}\bra{e^*} y_{e^*}+\ket{k^*}\bra{k^*} y_{k^*}\Bigl] \left( a^\dagger + a\right)
    \end{align} 
\end{subequations}
with $\Delta$ the coupling constant between the two excited states. Defining the oscillator position operator $x=\sqrt{\frac{1}{2m_\text{RC} \omega_{\text{RC}}}}\left(a^{\dagger} + a \right)$, the coupling constants $y_i$ can be related to the equilibrium RC positions in the different electronic states: $y_i = - \sqrt{\left( m_{\text{RC}}\omega_\text{RC}^3 \right) / 2} \times x^{\text{eq,}i}$ with $i \in [ \ket{e}, \ket{e^*}, \ket{k^*}]$. All numerics are performed in the electronic diabatic basis defined in the Hamiltonian above, however for the parameters we have chosen for the HBT (HBQ) system, an adiabatic (BO) representation leads directly to the effective barriered (barrierless) $S_1$ PESs that have been associated with these molecules (See Supplementary Information, SI).

As we shall see, dissipation of the RC dynamics due to `environmental’ degrees of freedom play a key and surprising role in explaining the unusual KIE in the Zhang model. Here, this environment is described by a bath of $800$ quantum vibration modes which are coupled to the proton wave packet in order to capture its damping and fluctuations induced by intramolecular and solvent vibrations. The vibrational environment Hamiltonian reads
\begin{subequations}
\renewcommand{\theequation}{\theparentequation.\arabic{equation}}
\begin{align}
            &H^\text{vib} = \sum\nolimits_{k} \omega_{k} \, b_k^{\dagger}b_k \\
             &H_\text{int}^\text{RC-vib} = \lambda_{\text{reorg}} \left(a^{\dagger} + a \right)^2 - \left(a^\dagger + a \right) \sum\nolimits_{k} g_{k} \left( b_k^{\dagger} +  b_k \right) 
\end{align} 
\end{subequations}
with $\omega_k$ and $b^{\left( \dagger \right)}$ the characteristic frequency and the annihilation (creation) operator of the $k^\text{th}$ vibration mode. Here the system-bath coupling is characterised by an Ohmic spectral density $J\left(\omega\right)=\sum_k g_k^2 \delta\left(\omega_k - \omega \right)$. Sometimes called a 'counter term', the reorganisation energy  $\lambda_{\text{reorg}} =   \int \frac{J(\omega)}{\omega}\text{d}\omega $ is also included explicitly, so as to remove any bath-induced changes of the primary RC potential that could interfere with our interpretation/analysis of the subsequent dynamics. When simulating deuterium, the mass $m_\text{RC}$ is multiplied by 2 with respect to the hydrogen case, while the corresponding frequency $\omega_\text{RC}$ is divided by $\sqrt 2$ \cite{scholes_ref_2023}. In addition, as the adiabatic potential energy surfaces (expressed as a function of position $x$) should be identical for H and D, this implies that the coupling factors $|g_{k}|^2$ and the corresponding spectral density $J(\omega)$ are divided by $\sqrt 2$ for D with respect to H. In a simple Redfield treatment of the  resulting (harmonic) relaxation rates for an Ohmic environment, we thus find that the relaxation rate for hydrogen is twice as large as that of deuterium (more details can be found in SI). These models are then simulated using a newly developed method within the Matrix Product State (MPS)- TEDOPA formalism \cite{chin_exact_2010, dunnett2021efficient} that allows to describe the many body wave function of the (real-space) proton wave packet alongside $all$ intramolecular vibrations on an equal, multi-dimensional footing \cite{lede_ESIPT_2024}. All the parameters of the models and parameters used to propagate the dynamics can be found in the Supporting Information (Table~S2-S4). Calculations were carried out with our open source MPSDynamics Julia package \cite{mpsdynamicsjl_2024,mpsdynamics_zenodo} which is available with full documentation and tutorials at \url{https://github.com/shareloqs/MPSDynamics}. \\

\begin{figure}[ht!]
        \centering
      \includegraphics[width=1.1\linewidth]{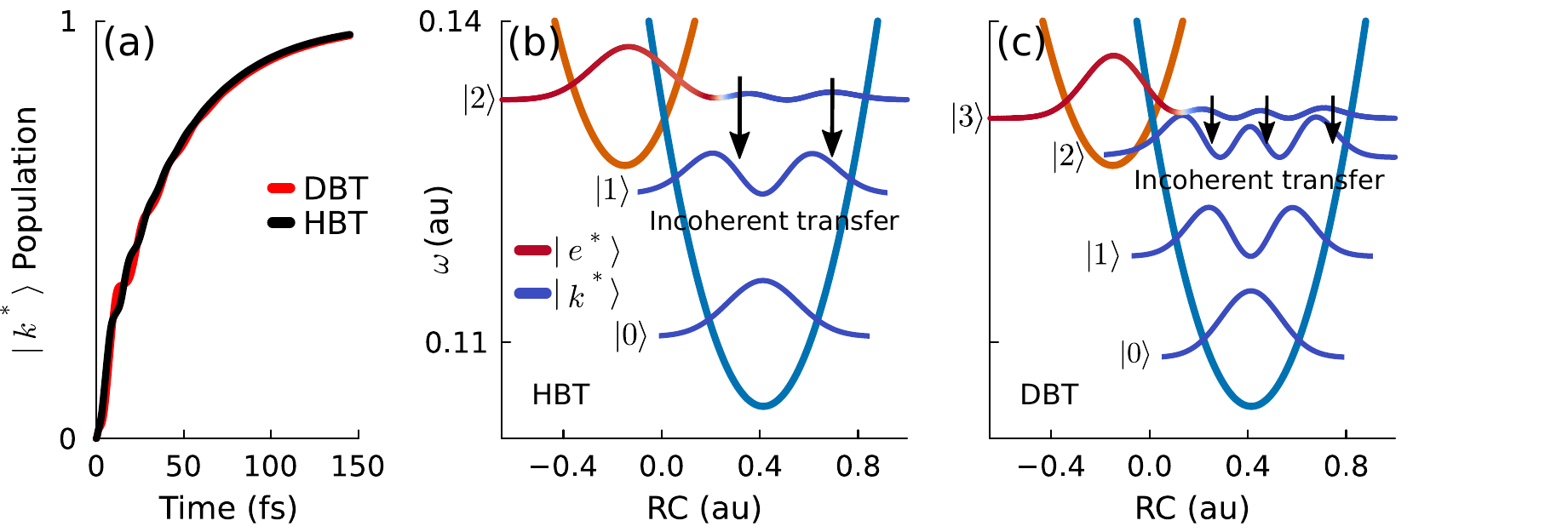}
      \caption{(a) $\ket{k^*}$ population dynamics for the Franck-Condon dynamics of HBT and DBT. (b) Eigenspectra of HBT, (c) and of DBT. Eigenstates are represented at their corresponding energy and the colorscale indicates their local electronic character of the states. Incoherent transfers for rate-determining steps are illustrated with vertical arrows.}
    \label{fig:ZPE_dynamics_BT}
\end{figure}

As a first step, dynamics are performed with the wave packet initialized according to the Franck-Condon principle, \textit{i.e.} in a RC state that corresponds to the ground vibrational state of the enol $\ket e$ instantaneously promoted to the enol excited state $\ket{e^*}$. This mimics an instantaneous optical excitation, i.e. the impulsive limit of extremely short pulses; the dramatic effect of finite pulse durations will be presented later. In Figure~\ref{fig:ZPE_dynamics_BT}~(a), for both HBT and DBT isotopes, the curves of the $\ket{k^*}$ population are almost $identical$, with a characteristic transfer time of $40~$fs that is well-described by a mono-exponential fit. Given the double-well shape of the adiabatic S1 PES, where proton tunnelling might be expected to play a dominant role at low temperatures, the absence of KIE is perhaps surprising, especially as we show in the SI that the ESIPT occurs almost entirely on the S1 surface.  However, this can be explained by inspecting the population transfers between the different $vibronic$ eigenstates of both systems (the first few states are shown in Figure~\ref{fig:ZPE_dynamics_BT}~(b) and (c), while more detailed eigenspectra in the diabatic and adiabatic bases can be found in SI). The detailed analysis provided in SI (section 3), shows that the rate-determining step in the overall ESIPT process is $\ket{2} \rightarrow \ket{1}$ for HBT and $\ket{3} \rightarrow \ket{2}$ for DBT. The absence of KIE in HBT/DBT essentially results from two compensating effects:  
(1) the damping rate towards lower energy states is two times slower in DBT than in HBT, whereas (2) the larger mass of D causes a near degeneracy of the 2nd keto vibronic state with the photo-excited enol state. Consequently, the ratio between the overlaps (transition matrix elements) of the relevant states involved in the rate-limiting transitions are found to obey  $||\braket{3|\hat x|2}||_\text{DBT}^2 = 2 \times ||\braket{2|\hat x|1}||_\text{HBT}^2$. As a consequence, the total decay rates that would be inferred from Fermi's Golden Rule have almost exactly the same value. 

This result is clearly model-dependent, although the $2 \times$ faster damping rate for H compared to D in equivalent environments should hold over the wide range of conditions where the common friction model we employ is valid. While the absence of a KIE in HBT (DBT) appears more naturally in the `passive' proton model, the current modelling highlights a general mechanism for how the same zero KIE can appear due to the alignment of product/reactant eigenstates in the presence of dissipation; indeed, it is the heavier mass of D, and the consequently greater density of 'accepting' states, that leads to its surprisingly fast ESIPT, emphasising the insight available in the open system perspective of this and many other ultrafast processes. This is particularly the case for the more contested mechanisms underlying ESIPT in HBQ (DBQ), which we now discuss.       

\begin{figure}[ht!]
        \centering
      \includegraphics[width=1.0\linewidth]{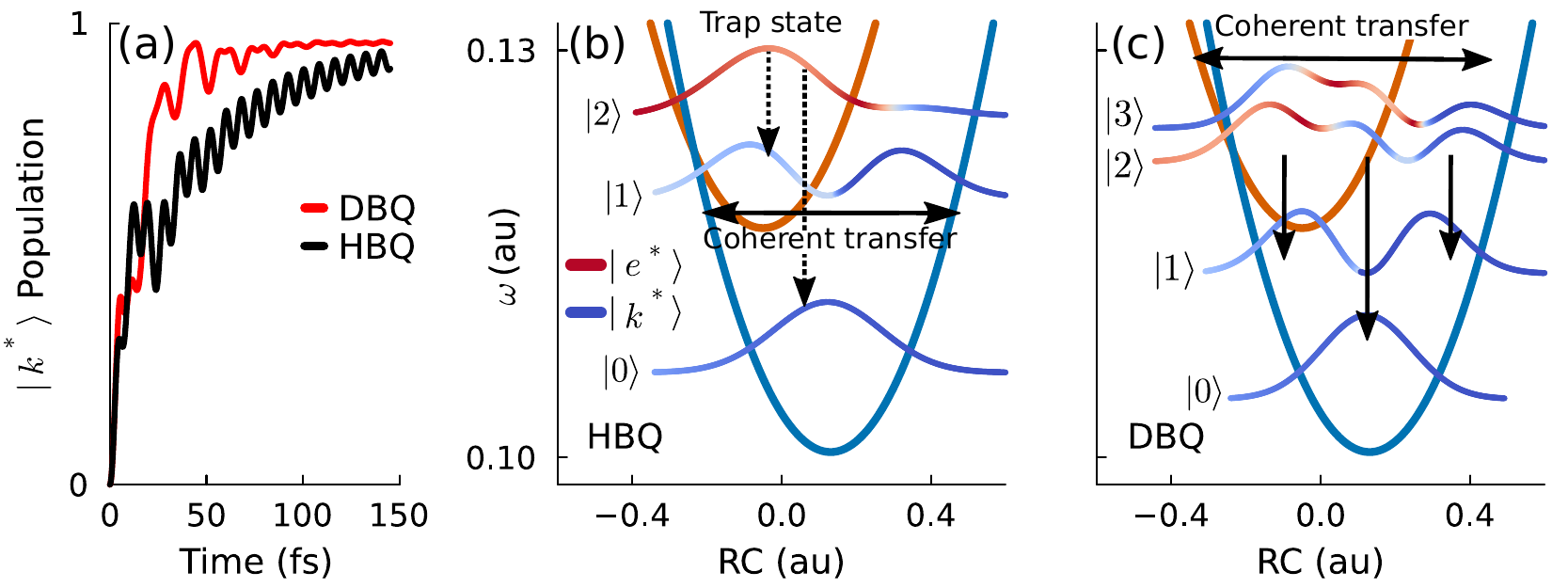}
      \caption{(a) $\ket{k^*}$ population dynamics for the Franck-Condon dynamics of HBQ and DBQ. (b) Eigenspectra of HBQ, (c) and of DBQ. Eigenstates are represented at their corresponding energy and the colorscale indicates their local electronic character. (Slow) incoherent transfers for rate-determining steps are illustrated with (dashed) vertical arrows and coherent transfers are represented by horizontal arrows.}
    \label{fig:ZPE_dynamics_BQ}
\end{figure}

For HBQ/DBQ, whose diabatic energy surfaces are reminiscent of an inverted Marcus configuration in electron transfer problems (Figure~\ref{fig:ZPE_dynamics_BQ}~(a)), both system exhibits a more complex and multiphasic behaviour.
In DBQ, the $\ket{k^*}$ population approaches unity yield in just a few tens of fs, whereas HBQ is only able to transfer $\approx 50 \%$ population to the keto excited state over the same timescale. After $\approx10$ fs, the population transfer in HBQ slows down significantly, showing weak, high-frequency oscillations on top of an exponential approach to unity ESIPT. 

In fact, both curves show high-frequency oscillations (faster in HBQ), that correspond to coherent oscillations between the diabatic $\ket{k^*}$ and $\ket{e^*}$ surfaces. In contrast to the case of HBT/DBT, which is driven $solely$ by incoherent dissipative processes, the seemingly irregular isotope effect, where proton transfer is significantly $faster$ for DBQ than for HBQ, originates from transient quantum coherence effects.  Figures~\ref{fig:ZPE_dynamics_BQ}~(b) and (c) show the eigenstate structure for HBQ and DBQ, respectively. Again, we note that the heavier DBQ presents a vibronic excited state of keto-character in near resonance with the impulsively photo excited enol state. Due to their strong coherent coupling -- and the effective absence of any barrier between adiabatic surfaces -- these states become strongly hybridised, allowing photoexcitation to excite a coherent superposition of the resulting eigenstates. This coherent wave packet then moves ballistically from reactant to product diabatic surfaces, showing oscillatory,  near-unity transfer between enol and keto excited states. Indeed this can be see in the absence of dissipation (See SI), where DBQ shows complete (reversible) oscillations between enol and keto states under purely unitary evolution. Further, the oscillation frequencies in the absence of friction correspond closely to those seen in Fig. \ref{fig:ZPE_dynamics_BQ} (c), matching the energy gaps of diabatic eigenstates involved in the initial wave packet (Figure S3 in SI and corresponding analysis). 

What is particularly striking in the present modelling is that the microscopic description of the dissipation for DBQ sets a timescale for decoherence/relaxation that is almost commensurate with a half period of the coherent oscillations: the coherent dynamics are thus able to transfer the proton with close to unity yield in a single half-oscillation, but the dissipation then prevents the reversible dynamics and effectively fixes the fast formation of the keto state. Again, while perhaps appearing somewhat fortuitous, this type of optimised interplay of coherent and dissipative effects has been proposed extensively in other ultrafast processes such as photosynthetic exciton energy transfer \cite{rather2024coherence,scholes2017using,chenu2015coherence,rather2018fundamental}.

In comparison, HBQ presents significant energy gaps between the initial enol and product keto eigenstates, so the initial wave packet is a weaker superposition of a delocalized product-reactant state and a highly localized product (enol-like) state ($\ket{2}$ in Fig.~\ref{fig:ZPE_dynamics_BQ}~(b)). In the absence of dissipation, this leads to periodic population oscillations that only transfer $\approx 50\%$ of the initial state to the keto ESIPT product. In the presence of dissipation, the delocalized component rapidly relaxes into the lowest keto state, but the `trapped' (localised) enol component can only relax to the keto minimum via incoherent energy loss to the environment (see Fig.~\ref{fig:ZPE_dynamics_BQ}~(a)). Due to poor electronic overlap of the `trap' state and the lower lying, weakly hybridised keto states, these transitions are slow, which explains the two timescales we observe in the HBQ population dynamics. Analysing this case in an adiabatic picture, we find that the `trap' state actually corresponds to a population on the upper adiabatic surface which, as mentioned above, is in a Marcus-inverted regime. Unlike the cases of HBT and DBT, the ultrafast dynamics of HBQ and DBQ under impulsive excitation are strongly non-adiabatic and populate both S1 and S2 surfaces (See SI).

\begin{figure}[ht!]
        \centering
      \includegraphics[width=1.0\linewidth]{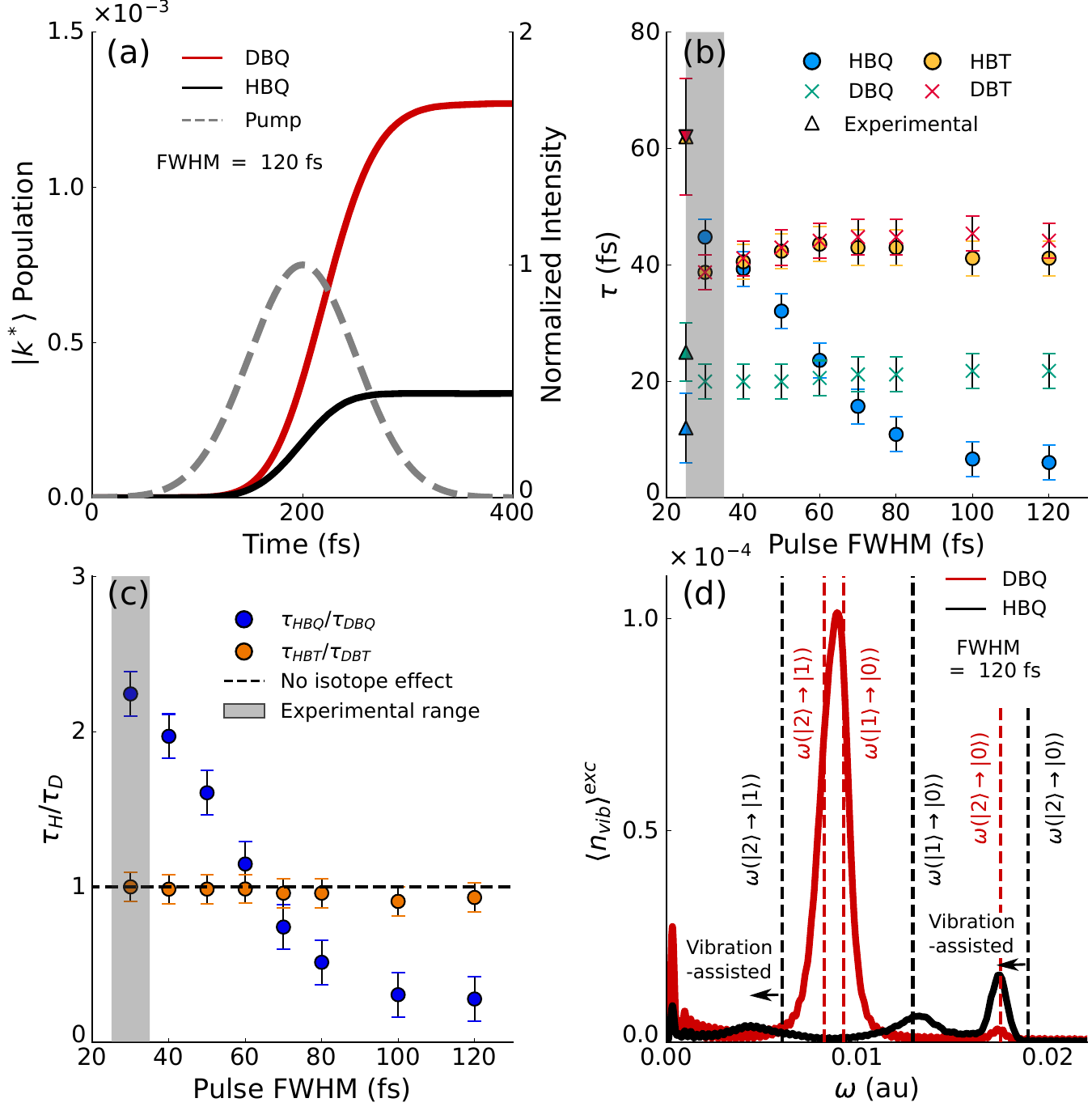}
      \caption{(a) $\ket{k^*}$ population dynamics promoted by a pulse of FWHM $=120~$fs for HBQ and DBQ. (b) Fitted transfer lifetime $\tau$ of the four systems for different pump FWHM. Experimental values of ref \cite{lee_active_2013} are also reported. (c) Isotope ratio of transfer lifetime for the different pulses indicating the KIE flip for HBQ and DBQ. Error bars indicate the fit uncertainty. (d) Vibrational spectra at final time of HBQ and DBQ promoted by a pulse of FWHM $=120~$fs.}
    \label{fig:isotope_effectv2}
\end{figure}

We now introduce pulsed excitation, by considering a wave packet initially in the ground $\ket e$ electronic state and promoted towards the excited $\ket{e^*}$ state through the time-dependent Hamiltonian:
\begin{subequations}
\renewcommand{\theequation}{\theparentequation.\arabic{equation}}
       \begin{align}
            &H^\text{drive}(t) =  \Big( \ket{e}\bra{e^*} + \ket{e^*}\bra{e}  \Big)\epsilon(t) \cos\big[ \left(\omega_{e^*}-\omega_{e} \right) t\big] \\
            &\epsilon(t) = \epsilon \exp{\left(-(t-t_0)^2/(2 \tau_\text{drive}^2)\right)}
    \end{align}
\end{subequations}
The pump pulse is centered at $t_0 = 200~$fs, with an amplitude $\epsilon $ and pulse duration $\tau_\text{drive}$. The temporal full width at half maximum (FWHM) of the pulse is given by $\left( 2 \sqrt{2 \log(2)} \right)\tau_\text{drive} $. We then vary the pulse duration and derive a typical ESIPT transfer time $\tau$ for each value of the FWHM 
via a fitting procedure detailed in SI. Figure~\ref{fig:isotope_effectv2}~(b) shows the transfer time of both isotopes and both molecules, excited by pulses with FWHM ranging from $30~$fs to $120~$fs. Pulses used experimentally in Refs.~\cite{kim_coherent_2009,lee_active_2013} are around $25~$fs long, and our results agree well with the experimental values, except for the striking case of HBQ which is analysed in detail below. 
Figure~\ref{fig:isotope_effectv2}~(b) shows that the transfer time $\tau$ is essentially independent of the FWHM for HBT/DBT, maintaining the absence of a KIE for all durations. This is also the case for DBQ, where the effect of the pulse width is negligible and, as observed with the Franck-Condon initial conditions, the ESIPT is very fast, $\tau \approx 20~$fs. 

In contrast, HBQ shows a dramatic dependence on the pulse FWHM, with the HBQ transfer time decreasing by an order of magnitude as the pulse length increases. Consequently, the isotopic ratio of transfer times (Figure~\ref{fig:isotope_effectv2}~(c)) -- a measure of KIE --  for HBQ and DBQ changes dramatically with pulse duration, as HBQ can perform ESIPT $faster$ than DBQ for long pulses, but is $slower$ for short/impulsive excitation (at least with the central pump frequency $\omega_{e^*}-\omega_{e}$ used in these simulations).

This striking behaviour can be explained by noting that the broader the pulse's temporal FWHM, the narrower its frequency spectrum becomes. In HBQ, short/impulsive excitation places $50\%$ of initial population in the trap state $\ket{2}$, as this has the largest effective transition dipole moment of all enol excited states. In contrast, for longer pulses, the narrowed pulse spectrum ceases to cover the direct ground state transition to  $\ket{2}$, as it has a central frequency midway between the optical transitions to $\ket{2}$ and $\ket{1}$ (Fig. \ref{fig:ZPE_dynamics_BQ}~(b)).  In the absence of dissipation, continuous wave excitation at this frequency would not be able to excite the enol system at all. However, the presence of the environment permits indirect vibration-assisted excitation that involves the emission of a phonon at an energy corresponding to the difference between the pulse central frequency and the $\ket{1}$ energy. This vibration-assisted excitation results in a faster proton transfer because it preferentially populates $\ket{1}$, which is  delocalised (hybridised) over the reactant-product surfaces while suppressing the population in the S2 enol trap state.  

This analysis is supported by the direct visualisation of environment observables which reports directly on the relative weights and speeds of the different dissipative pathways contributing to the ESIPT dynamics \cite{schroder2019tensor,riva2023thermal,lede_ESIPT_2024}. Figure~\ref{fig:isotope_effectv2}~(d) illustrates the vibrational excitation spectrum at the end of the simulation for HBQ/DBQ, with $\braket{n_\text{vib}}(\omega_k)=\braket{b_k^\dagger b_k}$ the average number of excitation and $\langle n_\text{vib} \rangle^\text{exc} = \langle n_\text{vib} \rangle \otimes \left(1- |e\rangle \langle e |\right)$ its projection onto the excited states (additional spectra for the other system and for different excitation conditions are illustrated Figures~S10 to S12). In DBQ, the main peaks match the transitions energies between vibronic eigenstates, which is consistent with the picture of a coherent wave packet initially transferring population, followed by relaxation mediated by the vibrational environment. 

In contrast, in HBQ, for long pulses (FWHM$=120$~fs), the environmental spectra corresponding to $\ket{2} \rightarrow \ket{0}$ and $\ket{2} \rightarrow \ket{1}$ transitions that would be seen under impulsive excitation are significantly redshifted. For large FWHM, direct excitation to $\ket{2}$ is strongly suppressed, and the only possible optical transition towards $\ket{1}$ (or $\ket{0}$) require vibrational assistance, with the energy surplus -- given by the detuning of the pulse frequency from the $\ket{1}$ or $\ket{0}$ states -- emitted into in the vibrational environment. 

This explains the position of the three main peaks in the vibrational HBQ emission spectrum of Figure~\ref{fig:isotope_effectv2}~(d), corresponding to the vibration-assisted excitation to $\ket{0}$ or $\ket{1}$, with a third peak that can be associated to the successive relaxation from $\ket{1}$ to $\ket{0}$ that is independent of excitation conditions. Further corroborating analysis of the variations of the environmental emission spectra as a function of the pulse characteristics can be found in the SI (Figures~S9 to S12). This interpretation also sheds light on the sensitivity of the transfer time in HBQ with the detuning of the pulse driving frequency, presented in Figure~S9. The SI presents the highly non-monotonic HBQ action spectra of ESIPT as a function of quasi-CW pulse frequency. This clearly shows accelerated ESIPT when the laser frequency is resonant with delocalised reactant/product states, and much slower ESIPT when centered on localized/trapped states on the S$_2$ adiabatic surfaces.


In conclusion, the dynamics and H/D isotope effect in two different ESIPT models were studied according to the recent non-adiabatic  model of Zhang \textit{et al}.~\cite{scholes_ref_2023}  The counter-intuitive isotope effects found in experiments are explained by the different super positions of electronic character in the vibronic eigenstates involved and the balance between coherent and dissipative (incoherent) transfer. Additionally, the isotope effect in the barrierless model depends strongly on the initial conditions induced by excitation conditions, where vibration-assisted absorption and enhanced ESIPT have been predicted. 

We believe that this study opens an interesting new route for studying multidimensional models of ESIPT, and could naturally be extended to  system-specific experimental conditions via vibrational spectral densities obtained by \textit{ab initio} calculations or experiments. Moreover, the open-system and many body lens that we have applied in our analysis could provide a potent framework for elucidating the primary (multidimensional) mechanisms for a wide range of ultrafast dissipative processes in Chemistry, Material Science and Physics.  

\begin{acknowledgement}
We thank iSiM (Initiative Sciences et ing\'enierie mol\'eculaires) 
 from the Alliance Sorbonne Universit\'e for funding. AWC acknowledges support from ANR projects ACCETP (ANR-19-CE24-0028) and Radpolimer (ANR-22-CE30-0033).

\end{acknowledgement}

\begin{suppinfo}

Detailed eigenspectra with $\ket{e^*}$ and $\ket{k^*}$ eigenstates as well as in the adiabatic basis. Comparison between the Franck-Condon dynamics for the isolated system, for the full system and for the driven dynamics. Time-resolved contributions for Franck-Condon and pulse promoted dynamics. 
Fit of $\tau$ details. Pulse detuning effect on HBQ and DBQ $\tau$. Vibrational spectra of HBT and DBT as well as for HBQ with different pulses. Time-resolved vibrational spectra. Calculation of damping difference between isotope. Model and dynamics propagation parameters. MP4 files of reduced density matrix for the Franck-Condon dynamics. 

\end{suppinfo}

\bibliography{biblio}

\end{document}



\section{Eigenspectra}
\begin{figure}[ht!]
        \centering
      \includegraphics[width=0.8\linewidth]{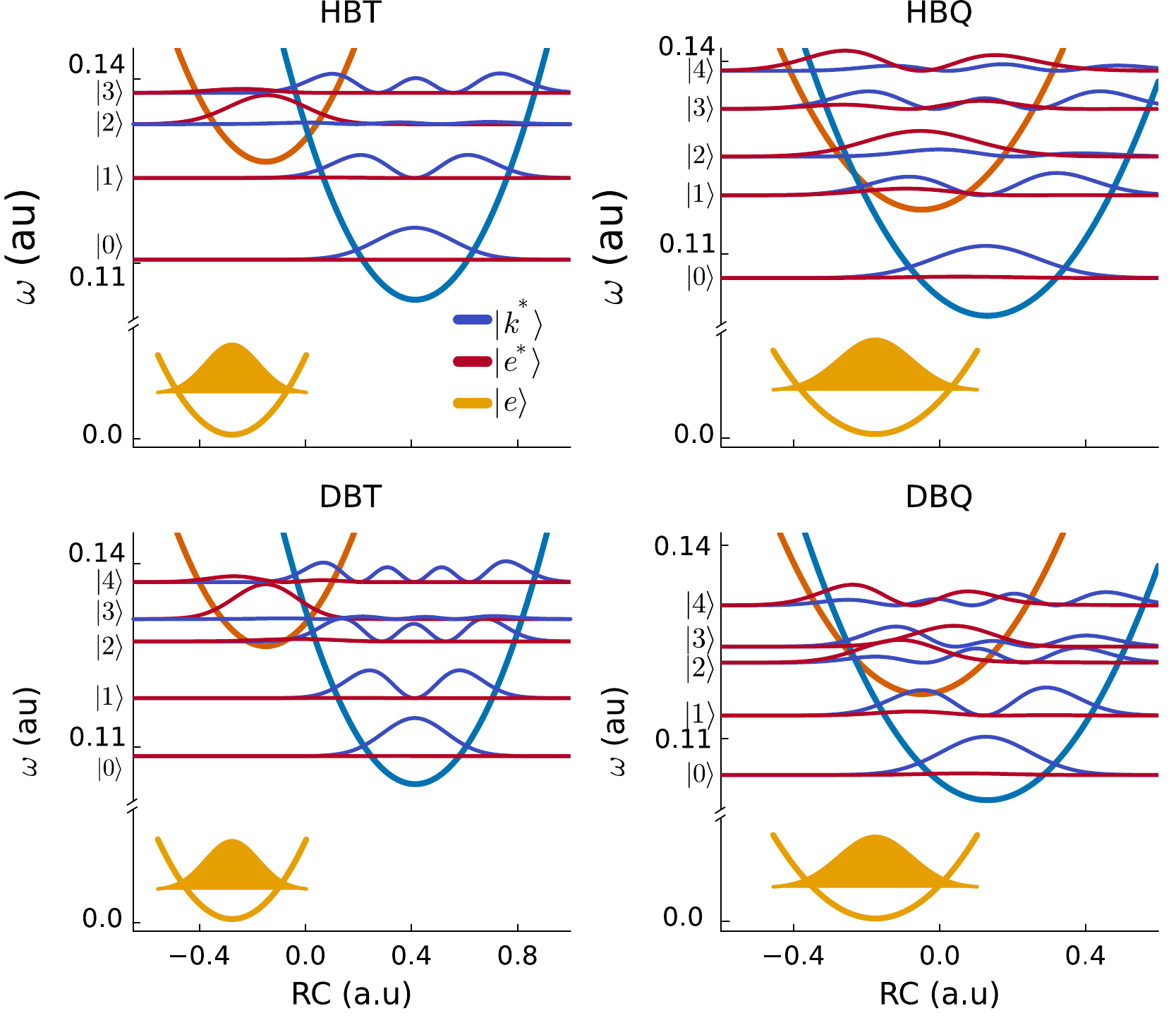}
    \caption{Eigenspectra in the diabatic basis of the four systems. The diabatic potential is represented with light colours. The shifted enol groundstate is also represented. }
    \label{figsup:ZPEdiab}
\end{figure}

\begin{figure}[ht]
        \centering
      \includegraphics[width=0.8\linewidth]{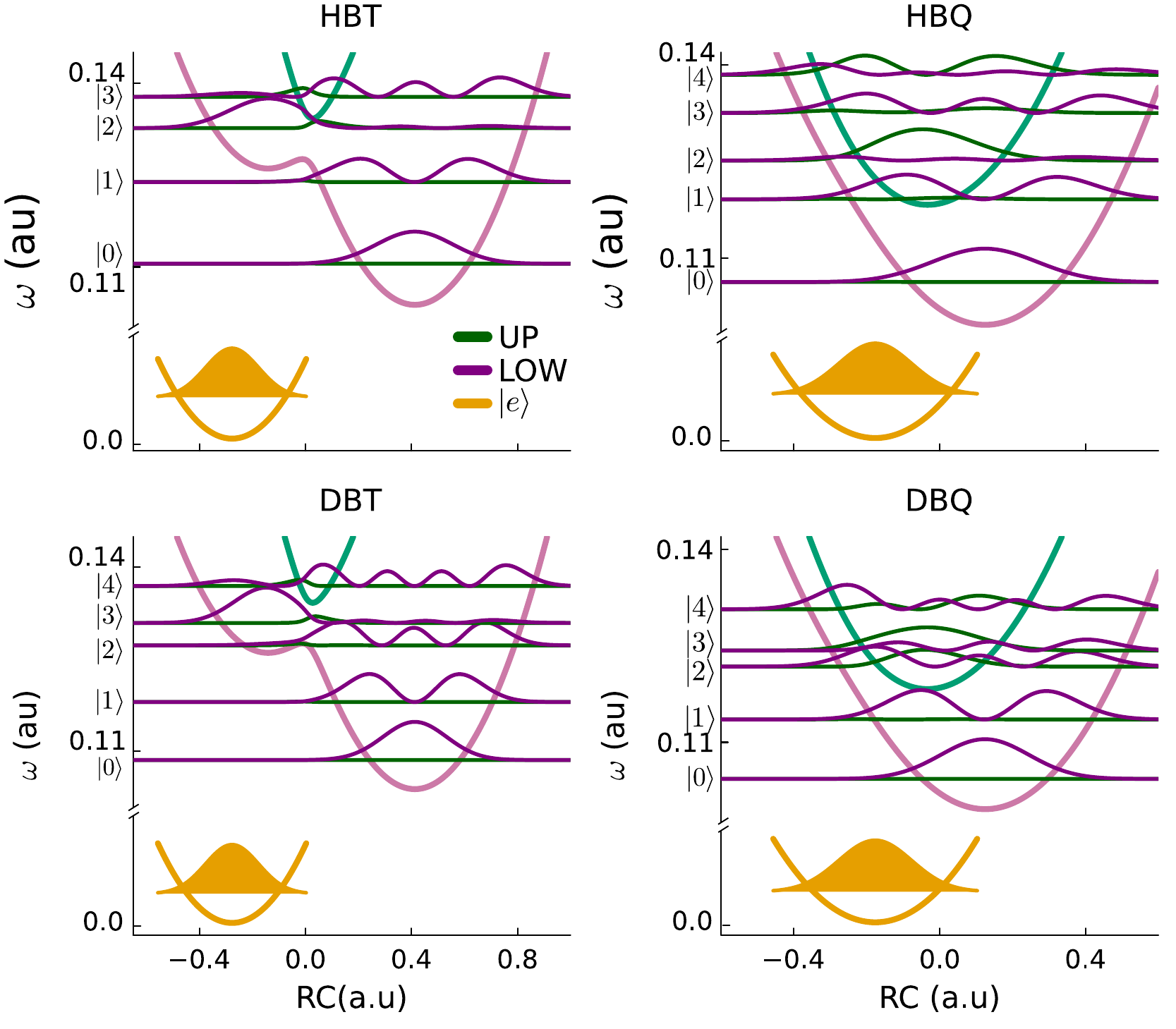}
    \caption{Eigenspectra in the adiabatic basis of the four systems. The adiabatic potential is represented with light colours. The shifted enol groundstate is also represented.}
        \label{figsup:ZPEadiab}
\end{figure}

Detailed eigenspectra of the isolated (electronic +RC) systems are illustrated in Figure~\ref{figsup:ZPEdiab} in the diabatic basis and in Figure~\ref{figsup:ZPEadiab} in the adiabatic basis. The adiabatic basis is obtained by diagonalization of the electronic Hamiltonian at each individual RC position $x$. In both figures, the eigenstates and eigenenergies are unchanged, only the eigenvector components are represented in one or the other basis.

\newpage

\section{Comparison between the isolated and open system dynamics and influence of the driving}

\begin{figure}[!ht]
    \centering
    \includegraphics[scale=0.45]{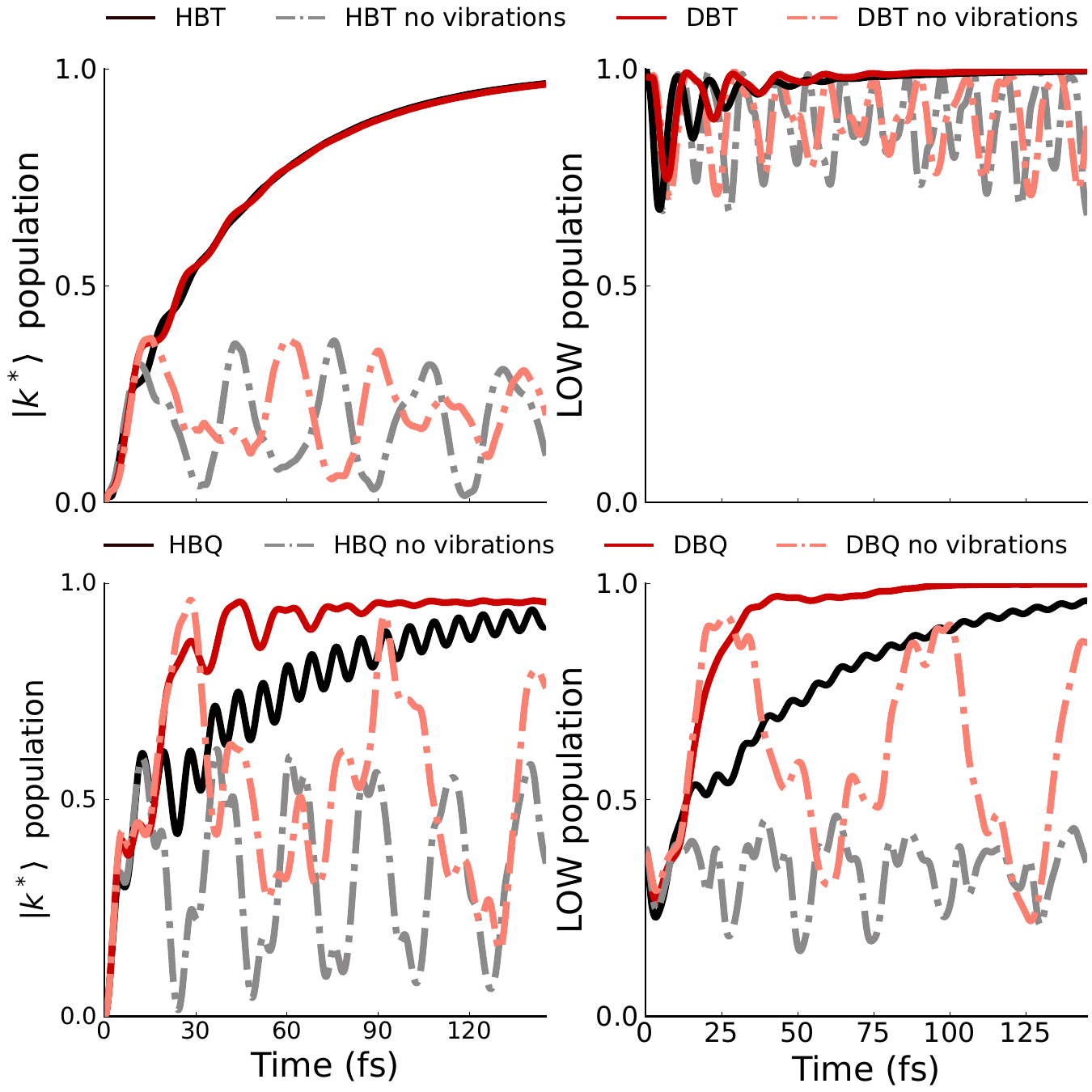}
           \caption{Comparison between the diabatic $\ket{k^*}$ (adiabatic LOW) population dynamics in left (right) figures between the isolated system in dashed lines and the system damped by vibrations in solid lines. This dynamics is initialised with the wavepacket in the Franck Condon region of $\ket{e^*}$.}
    \label{fig:Elecpop_BT}
\end{figure}

Figure~\ref{fig:Elecpop_BT} presents a comparison between the dynamics of the isolated system and that of the open system (coupled to the vibrational environment) as obtained with an initialisation according to the Franck-Condon principle. In all 4 cases (2 molecules and 2 isotopes), the two isolated and open system dynamics agree at very short times, before vibrations start having a significant influence on the wavepacket dynamics. Remarkably, the $\ket{k^*}$ population of DBQ undergoes an almost complete transfer even in the case of the fully coherent, isolated system dynamics.

\begin{figure}[ht!]
        \centering
    \includegraphics[width=0.7\linewidth]{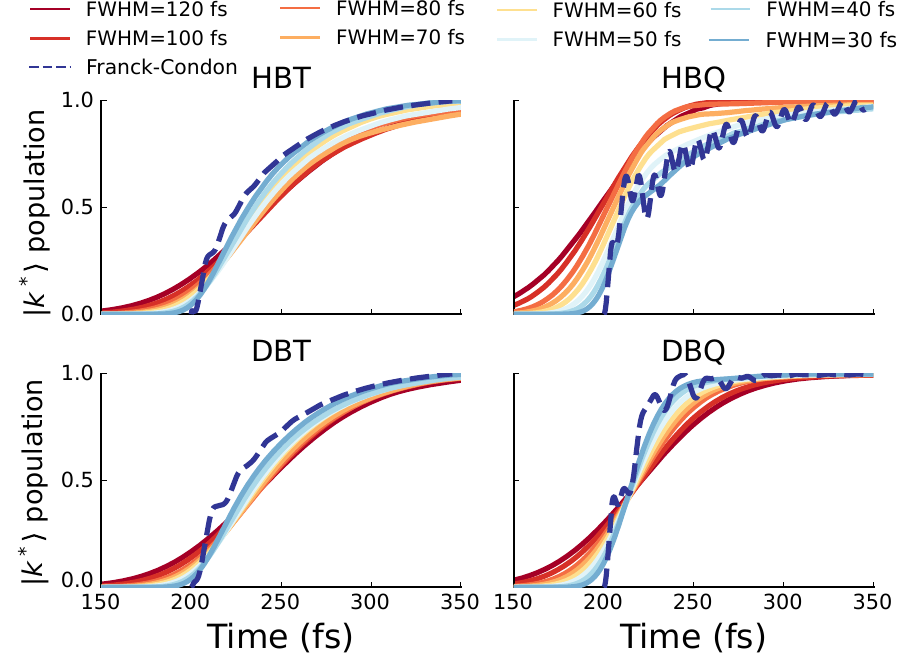}
    \caption{Normalized $\ket{k^*}$ population dynamics for different pulse durations and for the Franck-Condon dynamics. The Franck-Condon dynamics has been shifted to match the center of the pump pulses.}
     \label{figsup:kpop_FC_drive}
\end{figure}

In Figure~\ref{figsup:kpop_FC_drive}, the $\ket{k^*}$ (normalized) population curves are also compared between Franck-Condon dynamics and dynamics triggered by pump pulses of different durations. The shorter the pulse, the closer the populations approach the Franck-Condon results.

\newpage

\section{Dynamics of eigenstate contributions}

Population of the vibronic eigenstates (shown in Figure~\ref{figsup:ZPEdiab} and Figure~\ref{figsup:ZPEadiab}) can be probed over time. Table~S1 presents the contribution of each vibronic eigenstate $\ket{i}$ to the initial Franck-Condon state. These contributions arise from the displacement in equilibrium position $x^\text{eq}$ between the ground and the excited enol states $\ket{e}$ and $\ket{e^*}$.

    \begin{table}
    \caption{Initial eigenstate population for the Franck-Condon dynamics. This distribution stems from the shift along the RC between the $\ket{e}$ and $\ket{e^*}$ surfaces.}
    \centering
\begin{tabular}{|c|c|c|c|c|}
\hline
      $\,$  & HBT & DBT & HBQ & DBQ\\
   \hline    $\ket{0}$ & 1.55 $\times 10^{-4}$ & 1.54 $\times 10^{-5}$& 2.34 $\times 10^{-2}$& 1.54 $\times 10^{-2}$\\
      \hline  $\ket{1}$ & 6.30 $\times 10^{-3}$& 4.47 $\times 10^{-4}$& 0.152& 6.68 $\times 10^{-2}$\\
      \hline   $\ket{2}$ & 0.725 & 3.00 $\times 10^{-2}$& 0.654 & 0.475 \\
      \hline    $\ket{3}$ & 9.57 $\times 10^{-2}$&  0.677 & 8.04$\times 10^{-3}$& 0.203 \\
      \hline     $\ket{4}$ & 0.150 &  8.85 $\times 10^{-2}$& 0.143& 0.123 \\
      \hline     $\ket{5}$ & 3.23 $\times 10^{-4}$& 0.165&4.96$\times 10^{-4}$ & 8.26 $\times 10^{-2}$\\
      \hline       $\ket{6}$ & 1.83 $\times 10^{-2}$& 1.45 $\times 10^{-4}$& 1.65 $\times 10^{-2}$& 1.29 $\times 10^{-2}$\\
      \hline      $\ket{7}$ & 1.56 $\times 10^{-4}$& 3.13 $\times 10^{-2}$& 6.48 $\times 10^{-4}$& 1.72 $\times 10^{-2}$\\
      \hline    $\ket{8}$ & 1.65 $\times 10^{-3}$&3.01 $\times 10^{-6}$ & 1.30$\times 10^{-3}$& 4.39 $\times 10^{-4}$ \\
      \bottomrule
    \end{tabular}
\end{table}

\begin{figure}
    \centering
        \includegraphics[width=0.6\linewidth]{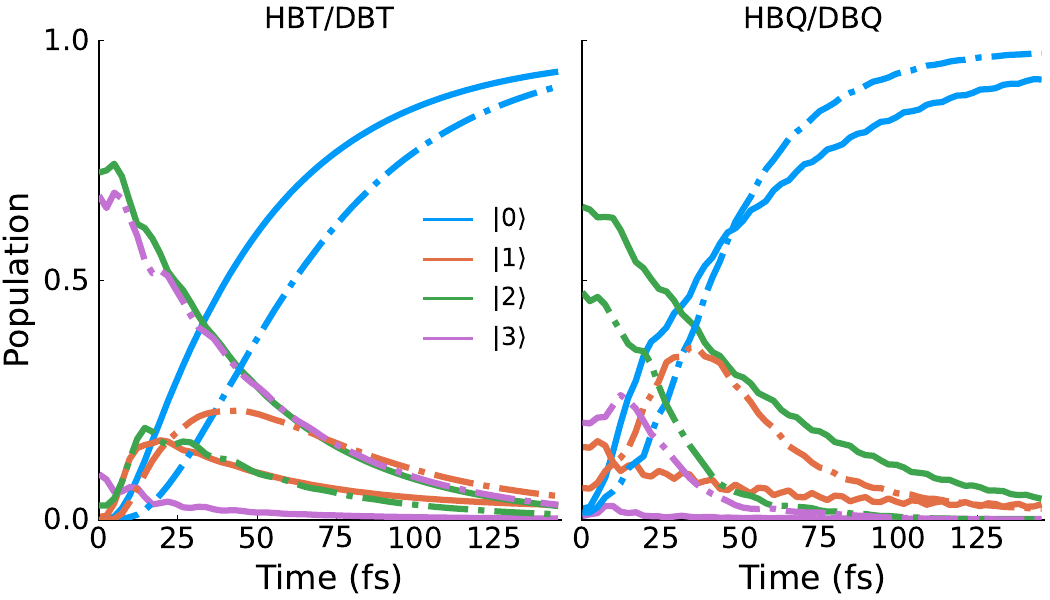}
\caption{Contribution of the lowest eigenstates for the Franck-Condon dynamics. The hydrogenated species are represented in solid lines while deuterated isotopes are in dotted-dashed lines.}
\label{figsup:contrib_nodriving}
\end{figure}

Figure~\ref{figsup:contrib_nodriving} shows the evolution in time of the vibronic eigenstates contributions for the Franck-Condon dynamics. The contribution of $\ket{2}$ for HBT and $\ket{3}$ for DBT follow the same trend, highlighting similar kinetic determinant step. Concerning DBQ, the $\ket{1}$ state is rapidly populated, illustrating the fast and coherent electronic transfer. 

\begin{figure}[!ht]
    \centering
    \includegraphics[scale=0.55]{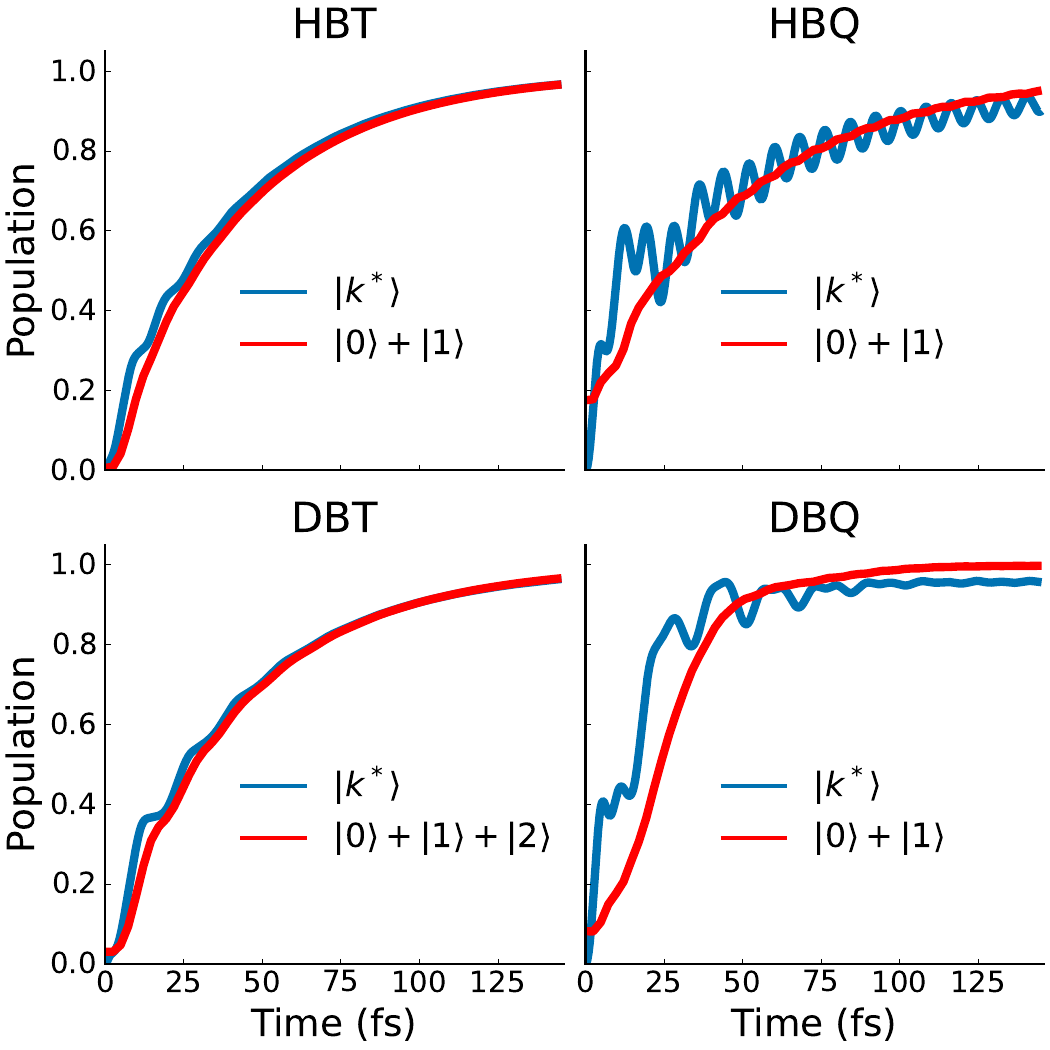}
    \caption{Comparison between the electronic population $\ket{k^*}$ and the low-energy eigenstate contributions for the Franck-Condon dynamics. The discrepancy for HBQ and DBQ at final time comes from the small $\ket{e^*}$ contribution of the $\ket{0}$ state.}
    \label{figsup:contrib_kcompar}
\end{figure}

Comparing with the total diabatic population $\ket{k^*}$, we identify the vibronic eigenstates  mainly involved in the electronic transfer process (Figure~\ref{figsup:contrib_kcompar}). In HBT (DBT), the total $\ket{k^*}$ population can be recovered by summing up the contributions of the purely $\ket{k^*}$ eigenstates $\ket 0+\ket 1$ (respectively $\ket 0+ \ket 1 +\ket 2$), whereas mixed states are involved for HBQ and DBQ, so that larger discrepancies are visible for these systems (even at the final simulation time, due to the small $\ket{e^*}$ contribution in the $\ket{0}$ state).

\newpage

\begin{figure}[ht!]
        \centering
    \includegraphics[width=0.8\linewidth]{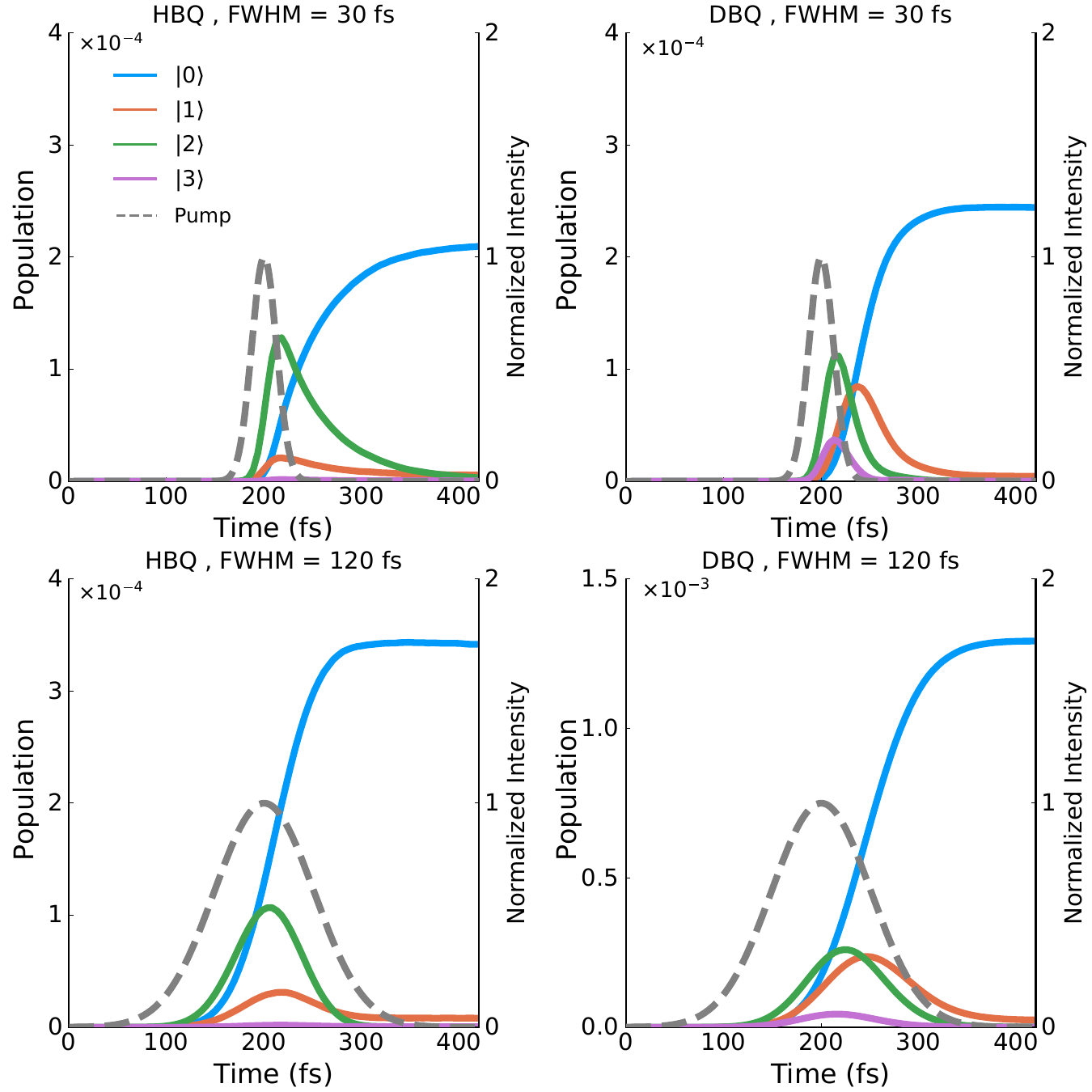}
    \caption{Contribution of the lowest eigenstates for the pulse initiated dynamics for HBQ and DBQ.}
     \label{figsup:contrib_driving}
\end{figure}

The same analysis can also be performed for dynamics initiated with a pump pulse. Here, we show in Figure~\ref{figsup:contrib_driving} the comparison between dynamics for short and long pulses in HBQ and DBQ. This comparison reveals a different eigenstate composition for HBQ between the different pulse widths.

\section{Transfer lifetime determination}
    
When driven, the dynamics of the system is assumed to approximately follow the simple kinetic equations:
\begin{align}
    &\partial_{t} P_{\ket{e^*}} = \alpha_{abs} \times I(t -(t_0 + \delta t)) - \frac{P_{\ket{e^*}}}{\tau} \\
    &\partial_{t} P_{\ket{k^*}} = \frac{P_{\ket{e^*}}}{\tau} \\
    &\partial_{t} P_{\ket{e}} = 1 - \alpha_{abs} \times  I(t -(t_0 + \delta t)) .
\end{align}
with $t_0$ the pulse central time, $\delta t$ the response delay of the environment, $\alpha_{abs}$ the absorption coefficient and $\tau$ the transfer rate.

\begin{figure}[ht!]
        \centering
    \includegraphics[width=0.8\linewidth]{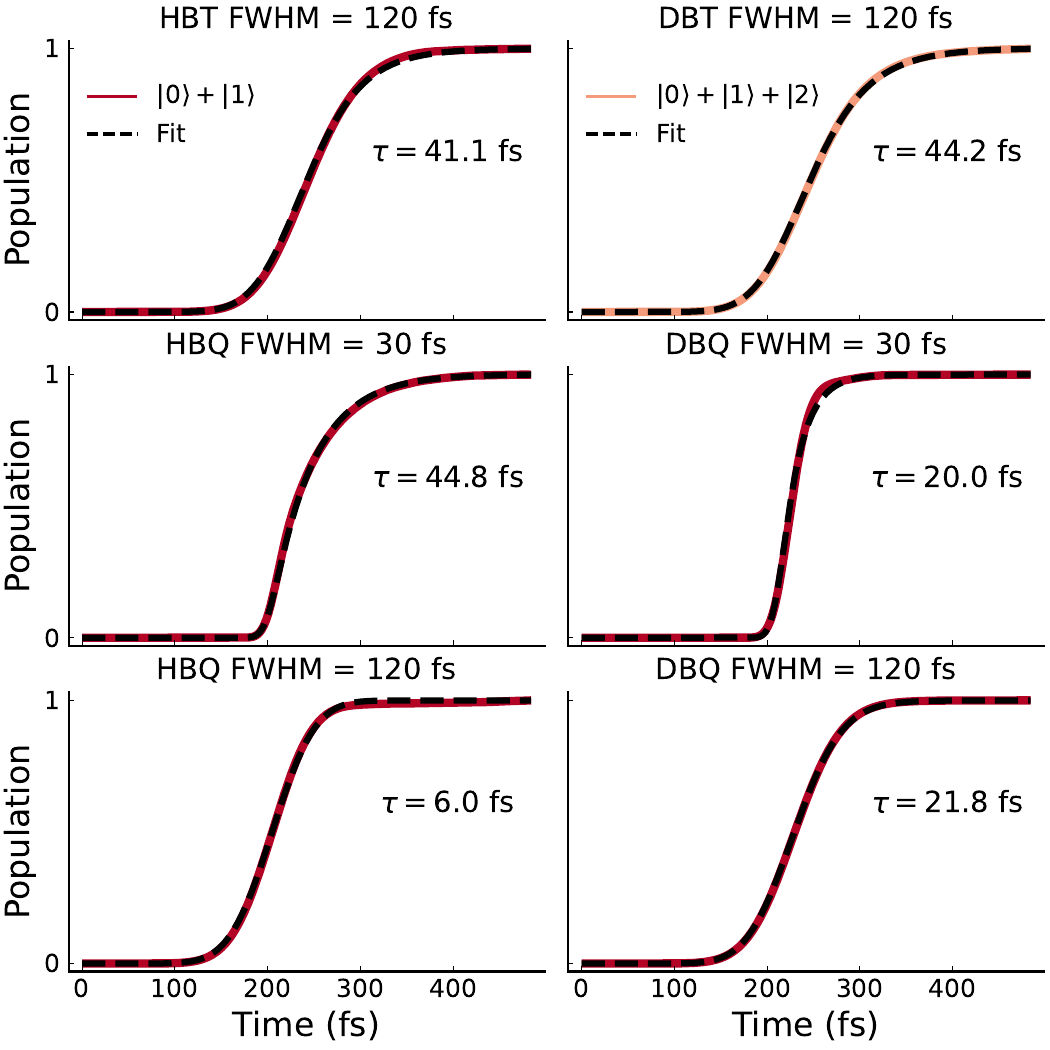}
    \caption{Fit on the normalized population of lower-lying eigenstate to determine the transfer lifetime $\tau$ for different pulses. We set $\delta t = 10~$fs for HBT, DBT and DBQ and $\delta t = 0~$fs for HBQ. }
    \label{figsup:fit}
\end{figure}

In order to determine the transfer lifetime, the excited keto population $\ket{k^*}$ is approximated by the sum of the lower-lying eigenstate contributions ($\ket{0} + \ket{1}$ for HBT, HBQ and DBQ and $\ket{0} + \ket{1} + \ket{2}$ for DBQ, see Figure~\ref{figsup:contrib_kcompar}). From the kinetic equations, the time-dependent populations $P_{e^*}(t) = \int_0^t  dt' \, \exp(-(t-t')/\tau) \times \alpha_{abs} \times  I(t' -(t_0 + \delta t))$ and $P_{k^*}(t) = \frac{1}{\tau} \int_0^t dt' P_{e^*}(t')$ can be recovered, with $I(t) = \epsilon(t)^2$. Normalizing these quantities in order to put aside the $\alpha_{abs}$ value and focus on the value of $\tau$, the eigenstate populations are then fitted to $P_{k^*}(t)$ (Figure~\ref{figsup:fit}).

\newpage

\section{Analysis of the pulse detuning for HBQ and DBQ}

\begin{figure}[ht!]
        \centering
    \includegraphics[width=0.6\linewidth]{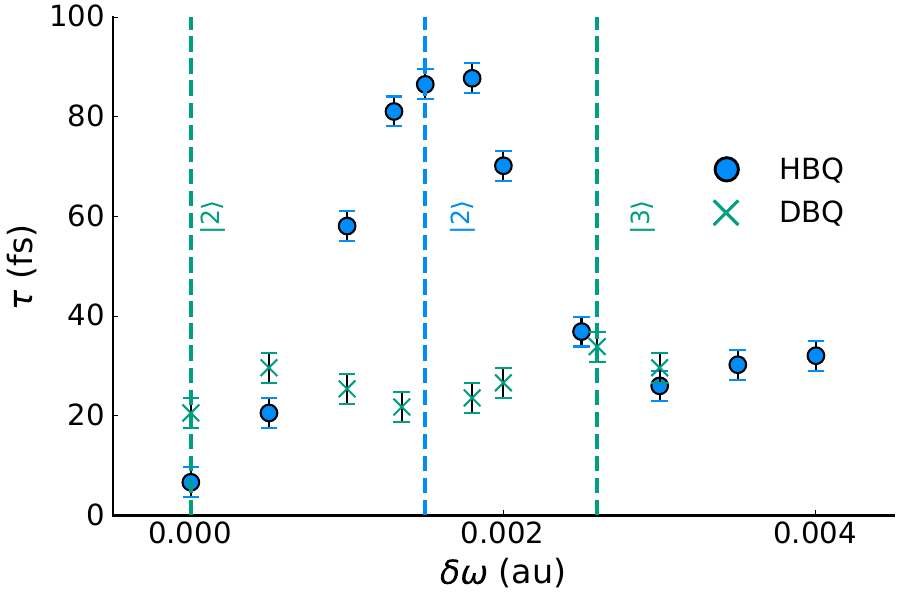}
    \caption{Transfer lifetime for different detuning value $\delta \omega$ of a pulse at FWHM = 120 fs for HBQ and DBQ. Error bars indicate the fit uncertainty. $\delta \omega = 0$ corresponds to the central frequency $\omega_{e^*}-\omega_{e}$ which is the frequency value used in the main text.}
    \label{figsup:detuning}
\end{figure}

For HBQ and DBQ, the isotope effect depends on the pulse width but also on its detuning with respect to the eigenenergies of the system. Due to the coupling with vibrations, even though a pulse is not stricly on resonance with one of the vibronic eigenstates, the ground state population can still be partially promoted to the excited state surface. By varying the detuning of a broad pulse (FWHM $= 120~$fs) with respect to the vibronic eigenenergies, a transfer pattern can be identified (Figure~\ref{figsup:detuning}). For HBQ, the transfer lifetime becomes longer when on resonance with the $\ket{2}$ trap state, whereas it is strongly reduced at zero detuning (corresponding to a central pulse frequency $\omega_{e^*}-\omega_{e}$, as in the main text). In contrast, the dependence on detuning is much weaker for DBQ.

\newpage

\section{Vibrational spectra}

\begin{figure}
        \centering
      \includegraphics[width=0.7\linewidth]{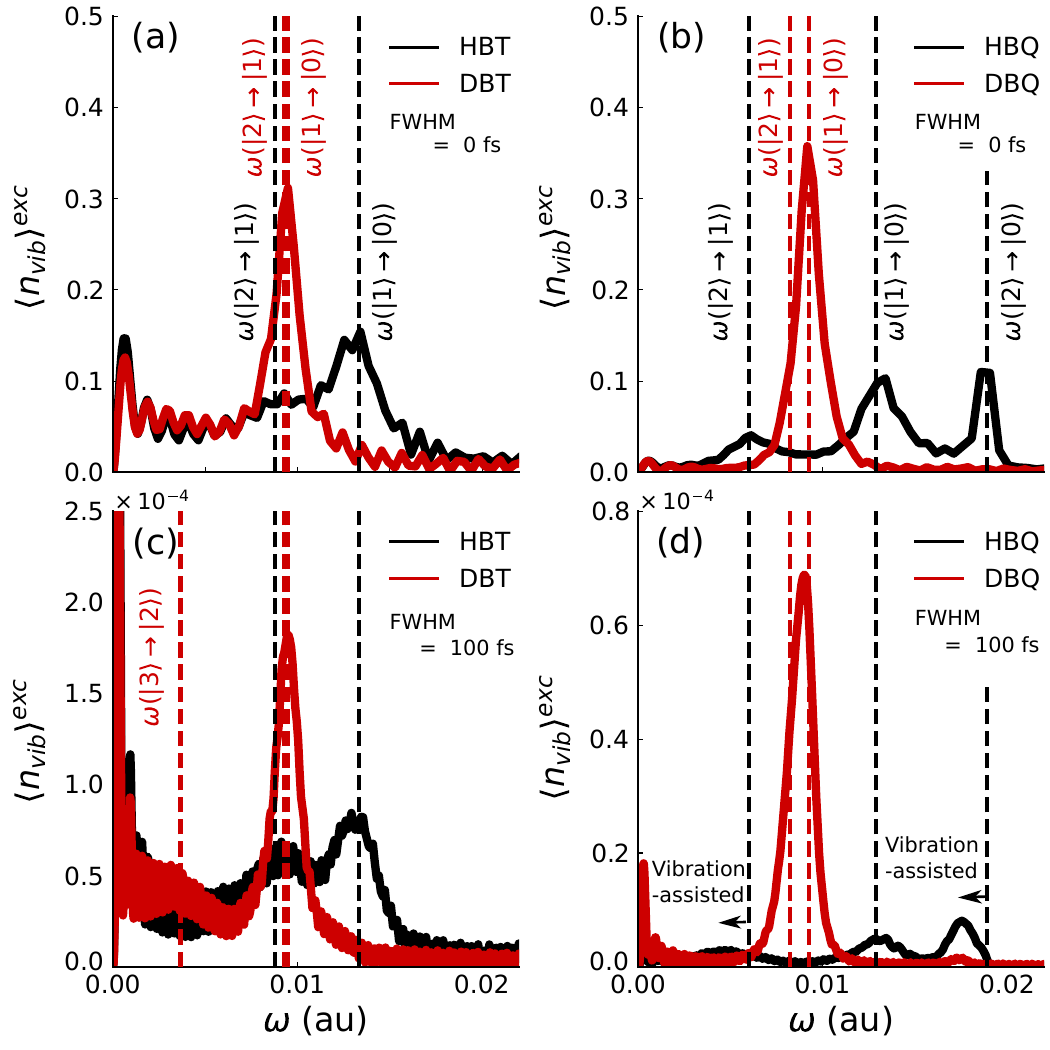}
      \caption{Vibrational spectra at final time (a)-(b) for the Franck-Condon initialisation and (c)-(d) for the driven system dynamics (FWHM = 100 fs). Peaks correspond to eigenstate gaps while the vibration-assisted absorption for HBQ shifts peaks involving the $\ket{2}$ state.}
    \label{figsup:vibspec_tot}
\end{figure}

Vibrational spectra of HBT, DBT, HBQ and DBQ are presented in Fig~\ref{figsup:vibspec_tot} for the Franck-Condon initialisation (FWHM = 0 fs) and for long pulses. Distinct peaks are probed and correspond to the different transition frequencies. While DBT and DBQ show a dominant peak that corresponds to the main transitions, the broad transition $\ket{2} \rightarrow \ket{1}$ for HBT is hidden under noises. For the driving case, HBT $\ket{2} \rightarrow \ket{1}$ peak is better resolved while the peak $\ket{3} \rightarrow \ket{2}$ for DBT is now visible.

\begin{figure}[ht!]
        \centering
    \includegraphics[width=0.8\linewidth]{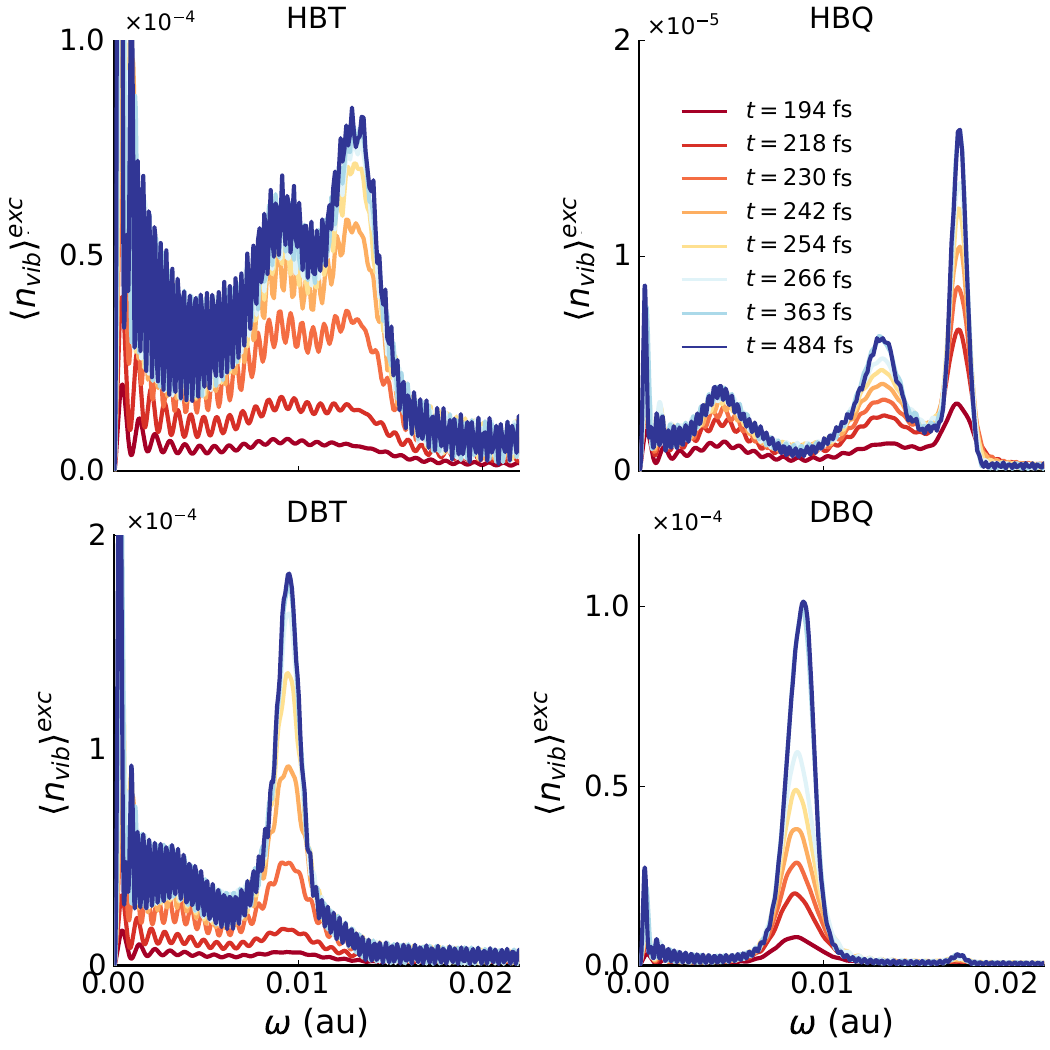}
    \caption{Vibrational spectra of HBT, DBT, HBQ and DBQ resolved in time for a dynamics initiated by a pulse (FWHM = 120 fs). }
    \label{figsup:phonspec_timeresolved}
\end{figure}

These spectra can also be probed at different time which gives insights into the dynamics. The example of HBQ and DBQ is presented in Figure~\ref{figsup:phonspec_timeresolved}. In particular for DBQ, it can be detected that at early time, the peak at energy $\omega(\ket{2} \rightarrow \ket{1})$ is first visible to be hindered by the peak at energy $\omega(\ket{1} \rightarrow \ket{0})$.

\begin{figure}[ht!]
        \centering
    \includegraphics[width=0.8\linewidth]{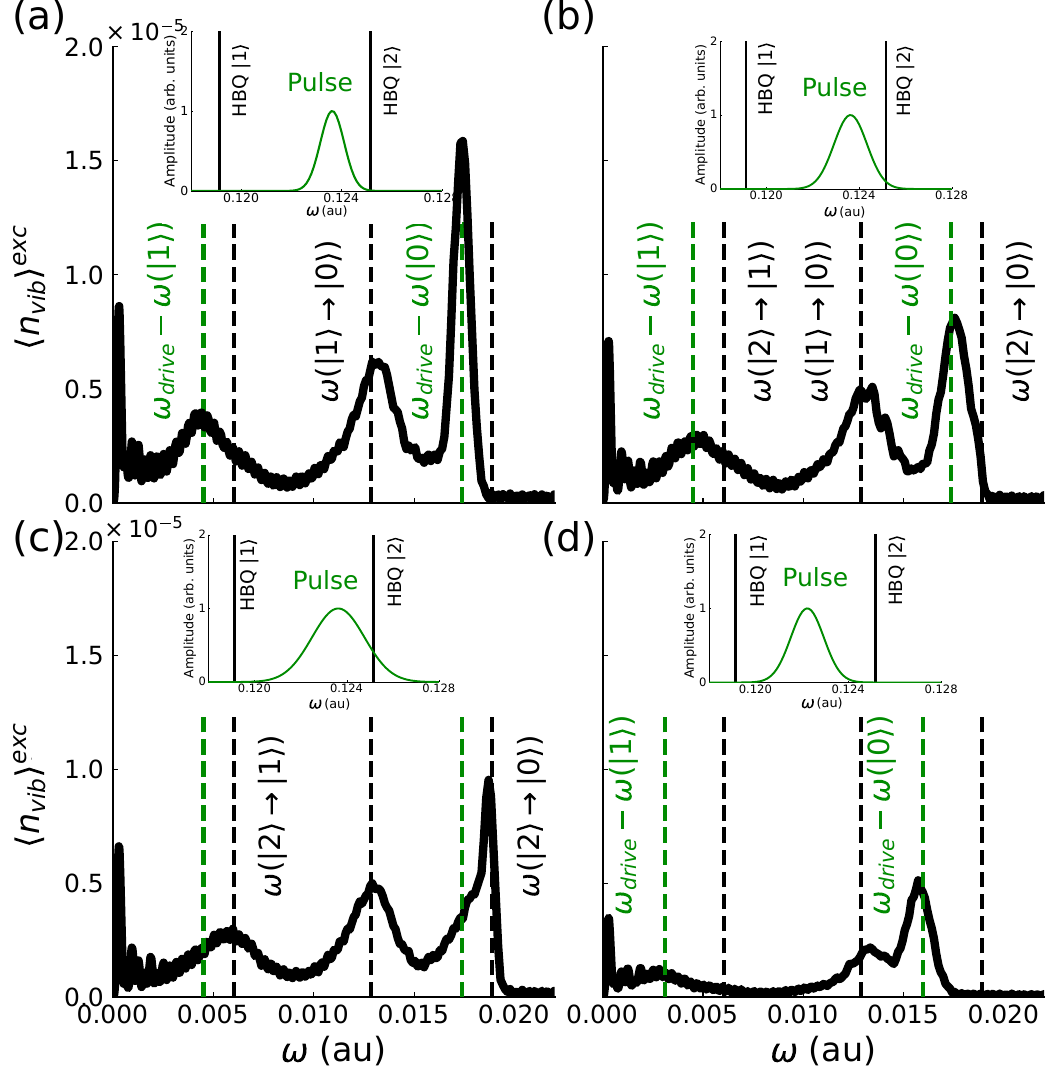}
    \caption{HBQ vibrational spectrum at final time for different pulses : (a) FWHM $= 120~$fs , $\delta \omega = 0~$au (b) FWHM $= 80~$fs , $\delta \omega = 0~$au (c) FWHM $= 50~$fs , $\delta \omega = 0~$au (d) FWHM $= 80~$fs , $\delta \omega = -0.0014~$au. Eigenstate energy gaps and vibration-assisted absorption are represented by vertical dashed lines. The Fourier Transform of the pulse is represented along eigenenergies of HBQ for each case.}
    \label{figsup:phonspec_HBQ_pulse}
\end{figure}

The vibrational spectrum of HBQ also undergoes changes when varying the pulse detuning or the pulse broadness as showed by Figure~\ref{figsup:phonspec_HBQ_pulse}. These vibrational spectra of HBQ are probed for different broadness at $\delta \omega = 0$ (FWHM = 120 fs, FWHM = 80 fs and FWHM = 50 fs). We clearly see that, the broader the pulse, the more vibration-assisted absorption the spectrum reveals. This phenomenon can be explained unambiguously with energy gaps and pulse envelope overlap with $\ket{2}$. The last example (Figure~\ref{figsup:phonspec_HBQ_pulse}~(d)) shows a pulse FWHM = 80 fs but detuned compared to the initial pulse. Peaks shift accordingly to the new $\omega_\text{drive}$.

\newpage

\section{Environment coupling of the isotopes}

The coupling of the RC oscillator to vibrational oscillator reads
\begin{subequations}
\renewcommand{\theequation}{\theparentequation.\arabic{equation}}
\begin{align}
       H_{\text{vib}} + H_{\text{int}}^{\text{RC-vib}} &= \left( \sum_k \frac{p_k^2}{2{m_k}} + \sum_k \frac{1}{2} {m_k} \omega_k^2 \left(q_k - \alpha_k \, x \right)^2\right)
\end{align}

with $q_k = \sqrt{\frac{1}{2{m_k}\omega_k}}\left(b_k + b_k^{\dagger} \right)$ the position operator of the $k^\text{th}$ mode of frequency $\omega_k$, {mass $m_k$} and $p_k = i \sqrt{\frac{{m_k}\omega_k}{2}}\left(b_k^{\dagger} - b_k \right)$ its momentum. Developing it in second quantization, it gives
\begin{align}
        H_{\text{vib}} &= \sum_k \omega_k b_k^{\dagger} b_k \\
       H_{\text{int}}^{\text{RC-vib}} &= \sum_k \frac{1}{2} {m_k} \omega_k^2 \left(-2 \, \alpha_k x q_k + \alpha_k^2 x^2\right) \\
          &= \sum_k \frac{\alpha_k^2 {m_k} \omega_k^2}{4m_\text{RC}\omega_\text{RC}} \left(a^{\dagger} + a \right)^2 - \left(a^{\dagger} + a \right) \sum_k \frac{\alpha_k \,  \omega_k \sqrt{{m_k} \, \omega_k}}{2\sqrt{m_\text{RC} \, \, \omega_\text{RC} }}  \left(b_k + b_k^{\dagger} \right) \label{eq:RC_vibcoupl} \\
             &= \lambda_{\text{reorg}} \left(a^{\dagger} + a \right)^2 - \left(a^{\dagger}  + a \right) \sum_k g_k \left(b_k + b_k^{\dagger} \right)  \label{eq:RC_vibcoupl2}
\end{align}
\end{subequations}

defining the coupling weights $g_k = \frac{\alpha_k \,  \omega_k \sqrt{{m_k} \, \omega_k}}{2\sqrt{m_\text{RC} \, \, \omega_\text{RC} }}$. The coupling amplitude between the reaction coordinate and the vibrational environment is interpreted with the spectral density $J(\omega) = \sum_k |g_k(\omega)|^2 \delta(\omega - \omega_k)$ that is taken as Ohmic with the continuous approach: $J(\omega) = 2 \alpha \omega \theta(\omega_c-\omega)$. Here, $\theta$ is the Heaviside function with $\omega_c$ the hard cutoff frequency. From Fermi's Golden Rule, the damping rate can be determined as :
\begin{equation}
    \gamma(H) = 2 \pi J^H(\omega_H)
\end{equation}
and $\gamma(H)$ is set up as $0.1~\text{fs}^{-1}$. To determine the damping rate of the deuterated isotope, keeping in mind that $m_D = 2 \, m_H $ and $\omega_D = \frac{\omega_H}{\sqrt{2}}$, one has from the coupling equations
\begin{align}
             &|g_k^H(\omega_k)|^2 = \left( \alpha_k^2 \omega_k^3 m \right)/ \left( 4 m_\text{H} \omega_\text{H} \right) \\
             & \frac{|g_k^H(\omega)|^2}{|g_k^D(\omega)|^2} = \frac{m_D \, \omega_D}{m_H \, \omega_H} = \sqrt{2} = \frac{J^H(\omega)}{J^D(\omega)}\\
             & \alpha_\text{H} / \alpha_\text{D} = \sqrt{2}
\end{align}
which gives
\begin{equation}
    \frac{\gamma(H)}{\gamma(D)} = \frac{J^H(\omega_H)}{J^D(\omega_D)} = \frac{\alpha_H \omega_H}{\alpha_D \omega_D} = 2
\end{equation}    

\section{Parameters}

The following parameters are used to simulate the HBT and HBQ systems. In the MPS formalism, we call the energy of the minimum of the well $\Tilde{\omega}_{i}$ with $i \in [ \ket{e}, \ket{e^*}, \ket{k^*}]$ which gives $\Tilde{\omega}_{i} = \omega_{i} - \frac{1}{2}m_\text{RC} \left(\omega_\text{RC} \, x_\text{RC}^{\text{eq,}i} \right)^2$ with ${\omega}_{i}$ defined in eq 1.2. Values are given in atomic units where $\hbar = m_e = e^2 =1$ and $m_\text{RC}=1.83 \times 10^3~$au for the proton.  \\

\begin{table} 
\caption{HBT and DBT system parameters. The value of the equilibrium displacement of the excited enol is arbitrarily chosen at $x_0^{e^*} = -0.15~$au.}
\centering
\begin{tabular}{|c|c|c|}
\hline
\textbf{Parameter} & \textbf{Value} & \textbf{Value (au)} \\
 \hline  $\hbar \Tilde{\omega}_{e^*}(x_\text{RC}^{\text{eq,}e^*})$ & $3.44 ~ \text{eV}$ & $0.1265~$\\
\hline $\hbar \Tilde{\omega}_{k^*}(x_\text{RC}^{\text{eq,}k^*})$ & $2.83 ~ \text{eV}$ & $0.1040~$\\
\hline $\hbar \omega_{\text{RC}}$(HBT) & $0.366 ~ \text{eV}$ & $0.01346~$\\
\hline $\hbar \omega_{\text{RC}}$(DBT) & $0.259 ~ \text{eV}$ & $0.00952~$\\
\hline $V \equiv \Delta$ & $0.1 ~ \text{eV}$ & $3.68 \, 10^{-3}~$\\
\hline $x_\text{RC}^{\text{eq,}e} - x_\text{RC}^{\text{eq,}e^*}$ & $0.63 $ & $0.127~$ $\equiv \gamma_{\text{displ}} $\\
\hline $x_\text{RC}^{\text{eq,}k^*}-x_\text{RC}^{\text{eq,}e^*}$ & $2.8$ & $0.564~$\\
\hline $\gamma_\text{mode}$ & $0.1 \, ~\text{fs}^{-1} $ & $\alpha_{\text{vib}}(H) = 1.43 \, 10^{-2}~$\\
\hline $\gamma_\text{mode}$ & $0.1 \, ~\text{fs}^{-1}$ & $\alpha_{\text{vib}}(D) = 1.01 \, 10^{-2}~$\\
\hline  $- \hat{\mu} \cdot \hat{\epsilon} \mu I \equiv \epsilon_{\text{drive}} $& $0.001 ~ \text{eV}$  & $ 3.68 \, 10^{-5}~$  \\
\hline  $\omega_{\text{drive}}$& $360 ~\text{nm}$  & $0.126~$ \\
\bottomrule
\end{tabular}

\end{table}

\begin{table}
\caption{HBQ and DBQ system parameters. The value of the equilibrium displacement of the excited enol is arbitrarily chosen at $x_0^{e^*} = -0.05~$au.}
    \centering
\begin{tabular}{|c|c|c|}
\hline
\textbf{Parameter} & \textbf{Value} & \textbf{Value (au)} \\
\hline  $\hbar \Tilde{\omega}_{e^*}(x_\text{RC}^{\text{eq,}e^*})$& $3.18 ~ \text{eV}$  & $0.1169~$  \\
\hline   $\hbar \Tilde{\omega}_{k^*}(x_\text{RC}^{\text{eq,}k^*})$& $2.73 ~ \text{eV}$  & $0.1004~$  \\
\hline  $\hbar \omega_{\text{RC}}$(HBQ)& $0.366 ~ \text{eV}$  & $0.01346~$  \\
\hline     $\hbar \omega_{\text{RC}}$(DBQ)& $0.259 ~ \text{eV}$  & $0.00952~$  \\
 \hline  $V \equiv \Delta $& $0.115 ~ \text{eV}$  & $4.23 \, 10^{-3}~$  \\
\hline     $x_\text{RC}^{\text{eq,}e} - x_\text{RC}^{\text{eq,}e^*}$& $ 0.63 $ & $0.127 \equiv \gamma_{\text{displ}}$   \\
\hline $x_\text{RC}^{\text{eq,}k^*}-x_\text{RC}^{\text{eq,}e^*}$ & $ 0.9 $  & $0.181~$  \\
 \hline   $\gamma_\text{mode} $& $0.1 \, ~\text{fs}^{-1} $ & $1.43 \, 10^{-2}~$   \\
 \hline   $\gamma_\text{mode} $& $0.1 \, ~\text{fs}^{-1} $ & $1.01 \, 10^{-2}~$   \\
\hline   $- \hat{\mu} \cdot \hat{\epsilon} \mu I \equiv \epsilon_{\text{drive}} $& $0.001 ~ \text{eV}$  & $ 3.68 \, 10^{-5}~$  \\
\hline    $\omega_{\text{drive}}$& $390 ~\text{nm}$  & $0.117~$ \\
        \bottomrule
    \end{tabular}
\end{table}

\begin{table}
                 \caption{Parameters simulations for the Franck-Condon (FC) dynamics (left) and for the driven system dynamics (right). Dynamics are carried out using the one-site Time-Dependent Variational Principle (1TDVP) \cite{haegeman_unifying_2016,paeckel_time-evolution_2019} and the rotating wave approximation.}

    \centering
\begin{tabular}{|c|c|c|}
\hline
\textbf{Dynamics Parameter} & \textbf{FC Value} & \textbf{Driven Value}   \\
\hline    $N_\text{vib}$& $190$ & $800$  \\
\hline    $d_\text{vib}$ & $10$ & $20$\\
\hline    $s_\text{vib}$ & $1.0$ & $1.0$  \\
\hline    $\alpha_\text{vib}$(HBT/HBQ) & $0.0143~$au & $0.0143~$au\\
\hline    $\alpha_\text{vib}$(DBT/DBQ) & $0.0101~$au & $0.0101~$au\\
\hline    $\omega_c$ & $5 \times \omega_{H/D}$ & $5 \times \omega_{H/D}$\\
\hline    $t_\text{final}$ & $6,000~$au & $20,000~$au\\
\hline    $dt$ & $2.5~$au & $2.5~$au \\
\hline    $D$ & $10$ & $16$\\
\hline    $d_\text{Fock}$ (HBQ/DBQ) & $25$  & $25$\\
\hline    $d_\text{Fock}$ (HBT) & $40$ & $45$\\
\hline    $d_\text{Fock}$ (DBT) & $30$ & $35$\\
        \bottomrule
    \end{tabular}
\end{table}

\bibliography{biblio}